\documentclass[conference]{IEEEtran}
\IEEEoverridecommandlockouts
\usepackage{cite}
\usepackage{amsmath,amssymb,amsfonts}
\usepackage{algorithmic}
\usepackage{graphicx}
\usepackage{color}
\usepackage{subfig}
\usepackage{stackengine}
\usepackage{comment}
\usepackage{ulem}
\usepackage{tabularx}
\usepackage{bm}
\usepackage{mathtools}
\usepackage{lettrine}
\usepackage{cuted}
\usepackage[colorlinks, citecolor=blue]{hyperref} % to highlight references with colors

\usepackage[flushleft]{threeparttable} % http://ctan.org/pkg/threeparttable
\usepackage{booktabs,caption}

\usepackage{textcomp}
\usepackage{xcolor}

\usepackage[linesnumbered, ruled]{algorithm2e}
\SetKwRepeat{Do}{do}{while}%

\SetCommentSty{mycommfont}

\def\BibTeX{{\rm B\kern-.05em{\sc i\kern-.025em b}\kern-.08em
    T\kern-.1667em\lower.7ex\hbox{E}\kern-.125emX}}
    
\newcommand{\noteblue}[1]{\textcolor{blue}{[{\bf #1}]}}

\newtheorem{problem}{\textit{Problem}}

\newtheorem {theorem*}{Theorem (Brouwer's Fixed Point Theorem)}

\usepackage{environ}
\NewEnviron{myequation}{%
\begin{equation}
\scalebox{1}{$\BODY$}
\end{equation}
}
 
\usepackage{xspace}
\usepackage{ifthen}
\usepackage[inline]{enumitem}

\newboolean{showcomments}
\setboolean{showcomments}{true}
\ifthenelse{\boolean{showcomments}}
{
}

\usepackage{subfig}
\usepackage{fancyhdr}

\newcommand*{\affmark}[1][*]{\textsuperscript{#1}}

\begin{document}

\title{Cooperate or not Cooperate:  Transfer Learning with Multi-Armed Bandit for Spatial Reuse in Wi-Fi}
%\author{\IEEEauthorblockN{ Pedro~Enrique~Iturria-Rivera and Melike~Erol-Kantarci, \IEEEmembership{Senior Member,~IEEE}}
%\IEEEauthorblockA{\textit{School of Electrical Engineering and Computer Science, University of Ottawa, Ottawa, Canada}} 
%Emails:\{pitur008, melike.erolkantarci\}@uottawa.ca}
\author{\IEEEauthorblockN{ Pedro Enrique Iturria-Rivera\affmark[1], \IEEEmembership{Student Member,~IEEE}, Marcel~Chenier\affmark[2], Bernard~Herscovici\affmark[2],\\ Burak~Kantarci\affmark[1], \IEEEmembership{Senior Member,~IEEE} 
and Melike Erol-Kantarci\affmark[1], \IEEEmembership{Senior Member,~IEEE}}
\IEEEauthorblockA{\affmark[1]\textit{School of Electrical Engineering and Computer Science, University of Ottawa, Ottawa, Canada}}  \affmark[2]\textit{NetExperience Inc., Ottawa, Canada}\\
Emails:\{pitur008, burak.kantarci, melike.erolkantarci\}@uottawa.ca,  \{marcel, bernard\}@netexperience.com  \vspace{-1em}}

%, Rick~Sommerville\affmark[2]
%rick.sommerville\} [2]

\maketitle
\begin{abstract}

\textbf{The exponential increase of wireless devices with highly demanding services such as streaming video, \iffalse Augmented Reality/Virtual Reality (AR/VR),\fi  gaming and others has imposed several challenges to Wireless Local Area Networks (WLANs). In the context of Wi-Fi, IEEE 802.11ax brings high-data rates in dense user deployments. Additionally, it comes with new flexible features in the physical layer as dynamic Clear-Channel-Assessment (CCA) threshold with the goal of improving spatial reuse (SR) in response to radio spectrum scarcity in dense scenarios. In this paper, we formulate the Transmission Power (TP) and CCA configuration problem with an objective of maximizing fairness and minimizing station starvation. We present four main contributions into distributed SR optimization using Multi-Agent Multi-Armed Bandits (MA-MABs). First, we propose to reduce the action space given the large cardinality of action combination of TP and CCA threshold values per Access Point (AP). Second, we present two deep Multi-Agent Contextual MABs (MA-CMABs), named Sample Average Uncertainty (SAU)-Coop and SAU-NonCoop as cooperative and non-cooperative versions to improve SR. In addition, we present an analysis whether cooperation is beneficial using MA-MABs solutions based on the $\epsilon$-greedy, Upper Bound Confidence (UCB) and Thompson techniques. Finally, we propose a deep  reinforcement transfer learning technique to improve adaptability in dynamic environments. 
Simulation results show that cooperation via SAU-Coop algorithm contributes to an improvement of 14.7\% in cumulative throughput, and 32.5\% improvement of PLR when compared with no cooperation approaches. Finally, under dynamic scenarios, transfer learning contributes to mitigation of service drops for at least 60\% of the total of users.}  

\end{abstract}

\section{Introduction}

\lettrine[findent=2pt]{\textbf{W}}{ }ireless connectivity has become an irreplaceable commodity in our modern society. The exponential trend expected in the wireless technology usage has lead analysts to predict that by 2023, $71\%$ of the global population will enjoy some kind of wireless service. In the group of Wireless Local Area Networks (WLANs), Wireless Fidelity (Wi-Fi) technology presents a growth up to 4-fold over a period of 5 years from 2018 to 2023 \cite{CiscoSystemsInc.2020}. The newest Wi-Fi standard IEEE-802.11ax \cite{9442429}, also known as Wi-Fi 6 expects to grow 4-fold by 2023 becoming $11\%$ of all the public Wi-Fi hostpots \cite{Cisco2020}.

Spatial reuse (SR) has been of interest for more than 20 years in the wireless community since it contributes to the reduction  of the collisions among stations and the determination of channel access rights \cite{Ye2003}. As the number of dense WLAN deployments increases, SR becomes more challenging in the context of Carrier Sense Multiple Access (CSMA) technology as used in  Wi-Fi\cite{Wilhelmi2021}. Wi-Fi 6 comes to address diverse challenges such as increasing number of Wi-Fi users, dense hotspots deployments and high demanded services such as Augmented, Mixed  and Virtual Reality. 

\begin{figure}[h]
\center
  \includegraphics[scale=0.24]{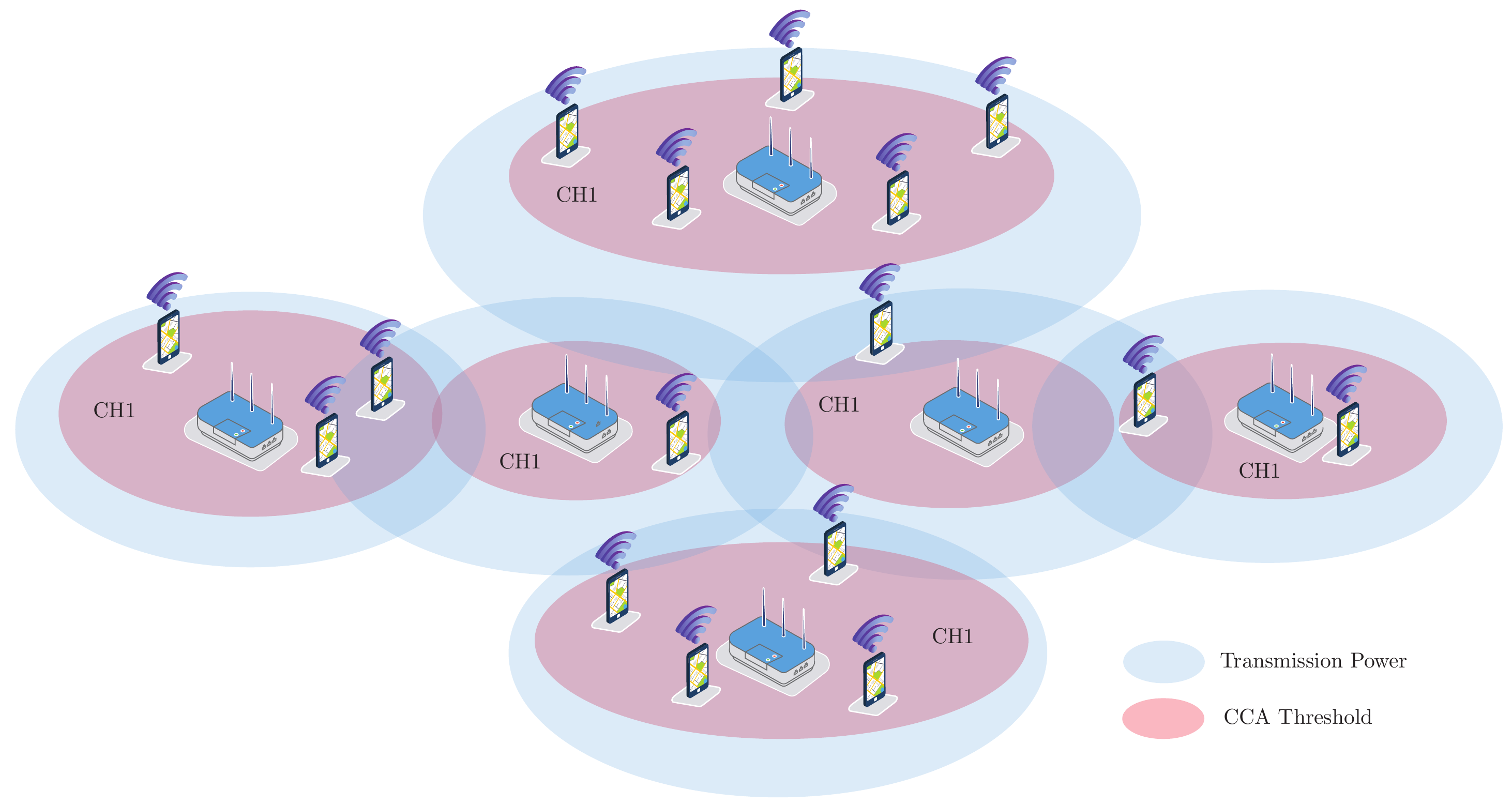}
  \caption{Typical operational scenario: APs adjust their Transmission Power and CCA threshold towards an efficient spatial reuse.}
  \label{state_vs_context}
%  \vspace{-2em}
\end{figure}

Moreover, 802.11ax included additional features such as dynamic adjustment of the Clear Channel Assessment (CCA) threshold and Transmission Power (TP). Static CCA threshold may not be representative of  diverse network topologies, and cause inefficient channel utilization or concurrent transmissions \cite{Thorpe2014}. Additionally, adjusting TP allows to reduce the interference among the APs and consequently maximize the network performance \cite{Huehn2012}. Thus, SR and network performance can be positively improved by adjusting CCA and TP. Yet, the complex interactions between CCA and TP, call for intelligent configuration of both. 

To this end, data scarcity and data access are key for any Machine Learning (ML) method \cite{Khosla2020}. Recently, AI-based wireless networks have been of remarkable interest among researchers both in WiFi domain \cite{szott2021wifi}, and 5G domain \cite{Elsayed2019} however the proposed solutions usually require complete availability of the data. In reality, data access is not always feasible due to privacy restrictions. Recent wireless network architectures have started to shift to a more open and flexible design. In 5G networks as well as the O-RAN Alliance architecture support the utilization of artificial intelligence to orchestrate main network functions \cite{iturria}. In the context of Wi-Fi, a novel project named OpenWiFi\cite{TIP2022} released by the Telcom Infra Project intends to disaggregate the Wi-Fi technology stack by utilizing open source software for the cloud controller and AP firmware operating system. These paradigm changes allow for the development of many applications in the area of ML and more specifically in Reinforcement Learning (RL) applications to become reality.

In this paper\footnote{The present work has been submitted to IEEE}, we intend to optimize TP and CCA threshold to improve SR and overall network KPIs.To do so, we formulate the TP and CCA configuration problem with an objective of maximizing product network fairness and minimizing station starvation. We model the SR problem as a distributed multi-agent decision making problem and use a Multi-Agent Multi-Armed Bandit (MA-MAB) approach to solve it. The contributions of this work, different from the ones found in the literature, can be summarized in the following points:
\begin{enumerate}
    \item We propose a solution for reducing the inherent huge action space given the possible combinations of TP and CCA threshold values per AP. We derive our solution via worst-case interference analysis.
    \item We analyse the performance of the network KPIs of well-known distributed MA-MAB implementations such as $\epsilon$-greedy, UCB and Thompson on the selection of the TP and CCA values in cooperative and non-cooperative settings.
    \item We introduce a contextual MA-MAB (MA-CMAB) named Sample Average Uncertainty-Sampling (SAU) in cooperative and non-cooperative settings. SAU-MAB is based on a deep Contextual MAB. 
    \item We propose for the first time, to the best of our knowledge, a deep transfer learning solution to adapt efficiently TP and CCA parameters in dynamic scenarios.
    
\end{enumerate}

With these contributions, our simulation results show that the $\epsilon$-greedy MAB solution improves the throughput at least 44.4\%, provides improvement of 12.2\% in terms of fairness and 94.5\% in terms of Packet Loss Ratio (PLR) over typical configurations when a reduced set of actions is known. Additionally, we show that the SAU-Coop algorithm improves the throughput by 14.7\% and PLR 32.5\% when compared with non cooperative approaches with full set of actions. Moreover, our proposed transfer learning based approach reduces the service drops by at least 60\%.

The rest of the paper is organized as follows. Section \ref{Section2} presents a summary of recent work that uses Machine Learning to improve SR in Wi-Fi. Section \ref{Section3} covers the basics on Multi-Armed Bandits including deep contextual bandits and deep transfer reinforcement learning. In \ref{Section4} we present our system model altogether with an analysis to reduce the action space via worst-case interference. Section \ref{Section5} presents the proposed schemes and the results are discussed in section \ref{Section6}. Finally, section \ref{Section8} concludes the paper.

%%%%%%%%%%%%
\section{Related work} \label{Section2}
Reinforcement learning-based spatial reuse has been of interest in recent literature. The studies have focused on distributed solutions with no cooperation or centralized schemes of multi-armed bandits. These studies are summarized below.

In \cite{Wilhelmi2019}, the authors present a comparison among well-known MABs as $\epsilon$-greedy, UCB, Exp3 and Thompson sampling in the context of decentralized SR via Dynamic Channel Allocation (DCA) and Transmission Power Control (TPC) in WLANs. The results showed that ``selfish learning'' in a sequential matter present better performance than ``concurrent learning'' among the agents. 
Additionally, \cite{Bardou2021} presents a centralized MAB consisting of an optimizer based on a modified Thompson Sampling (TS) algorithm and a sampler based on Gaussian Mixture (GM) algorithm to improve SR in 802.11ax Wi-Fi. More specifically, the authors propose to deal with the large action space comprised by TP and Overlapping BSS/Preamble-Detection (OBSS/PD) threshold by utilizing a MAB variant called Infinitely Many-Armed Bandit (IMAB). Furthermore, a distributed solution based on Bayesian optimizations of Gaussian processes to improve SR is proposed in \cite{bardou2022inspire}. 

Other solutions that are not related to reinforcement learning can be found in the literature with the aim of improving SR in WLANs. For instance, in \cite{9417353} the authors propose a distributed algorithm where the APs decide their Transmission Power based on their RSSI. Moreover, in \cite{app112211074} the authors present an algorithm to improve SR by utilizing diverse metrics such as SINR, proximity information, RSSI and BSS color and compare with the legacy existing algorithms. The ultimate goal of the previous algorithm is the selection of the channel state (IDLE of BUSY) at the moment of an incoming frame given the previous metrics. Finally, the authors in \cite{wilhelmi2022federated} presented a supervised federated learning approach for SR optimization. 

In all above works, the authors employ either centralized  or decentralized schemes with no cooperation to address SR optimization in WiFi. In this work, we propose to address this via a coordination based MA-MAB. In addition, we tackle some of the issues previously encountered in others works such as the size of action space due the set of possible values TP and CCA. Finally, to the best of our knowledge, we propose for the first time to address SR adaptation in dynamic environments utilizing deep transfer learning. 

\section{Background } \label{Section3}
In this section, we present a background on Multi-Armed Bandits including $\epsilon$-greedy, Upper Confident Bound, Thompson sampling bandits and an introduction on contextual MABs with a focus on a neural network-based contextual bandit. Additionally, we introduce MABs to the multi-agent setting and we finalize with a background on deep transfer reinforcement learning.    

Multi-Armed Bandits (MABs) are a widely used RL approach that tackles the exploration-exploitation trade-off problem. Their implementation is usually simpler when compared with full RL off-policy or on-policy algorithms. However, simplicity often comes with a cost of obtaining suboptimal solutions \cite{Bouneffouf2020}. The basic model of MABs corresponds to the stochastic bandit, where the agent has $K$ possible actions to choose, called arms, and receive certain reward $R$ as a consequence of pulling the $j^{th}$ arm over $T$ environment steps. The rewards can be modeled as independent and identically distributed (i.i.d), adversarial, constrained adversary or random-process rewards \cite{Slivkins2019}. From the four models previously mentioned, two are more commonly found in the literature: the i.i.d and the adversarial models. In the i.i.d model, each pulled arm's reward is drawn independently from a fixed but unknown distribution $D_j$ with an unknown mean $\mu_j^*$. On the other hand, in the adversarial model each pulled arm's reward is randomly sampled from an adversary or alien to the agent (such as the environment) and not necessarily sampled from any distribution\cite{zhou2015survey}. The performance of MABs is measured in terms of cumulative regret $R_T$ or total expected regret over the $T$ steps defined as: 
\begin{myequation}
    R_T = \sum_{t=1}^{T} \mathbb{E}[(\text{max}_j\mu_j^* - \mu_j^*)],
\end{myequation}
The utmost goal of the agent is to minimize $R_T$ over the T steps such as the $\lim_{T \to \infty} R_T/T$ = 0 which means the agent will identify the action with the highest reward in such limit. 

\subsection{$\epsilon$-greedy, Upper-Confidence-Bound and Thompson Sampling MAB}
The $\epsilon$-greedy MAB is one of the simplest MABs and as the name suggests, it is based on the $\epsilon$-greedy policy. In this method, the agent selects greedily the best arm most of time and once a while, with a predefined small probability ( $\epsilon$), it selects a random arm \cite{sutton2018reinforcement}. 
The UCB MAB tackles some the disadvantages of the $\epsilon$-greedy policy at the moment of selecting non-greedy arms. Instead of drawing randomly an arm, the UCB policy measures how promising non-greedy arms are close from optimal. In addition, it takes in to consideration the rewards' uncertainty in the selection process. The selected arm is obtained by drawing the action from $\text{\texttt{argmax}}_a\left[Q_{t}(a) + c\sqrt{\text{ln }{t}/N_{t}(a)}\right]$, where $N_{t}(a)$ corresponds to the number of times that action $a$ via the $j^{th}$ arm has been chosen and $Q_{t}(a)$ the Q-value of action $a$\cite{sutton2018reinforcement,Agrawal1995}. Finally, Thompson Sampling MAB action selection is based on Thompson Sampling algorithm as the name indicates. Thompson sampling or posterior sampling is a Bayesian algorithm that constantly constructs and updates the distribution of the observed rewards given a previously selected action. This allows the MAB to select arms based on the probability of how optimal the chosen arm is. The parameters of the distribution are updated depending on the selection of the distribution class\cite{Russo2018}.

\subsection{Deep Contextual Multi-Armed Bandits}
Contextual Multi-Armed Bandits (CMABs) are a variant of MABs, that before selecting an arm, observe a series of features commonly named context\cite{Bouneffouf2020}. \iffalse Figure \ref{state_vs_context}, depicts the  difference between the stateless MAB and CMAB. \fi Different from the stateless MAB, a CMAB is expected to relate the observed context with the feedback or reward gathered from the environment in $T$ episodes and consequently predict the best arm given the received features \cite{zhou2015survey}. Diverse CMABs have been proposed throughout the literature such as LinUCB, Neural Bandit, Contextual Thompson Sampling and Active Thompson Sampling \cite{Bouneffouf2020}. More recently, a deep neural contextual bandit named SAU-Sampling has been presented in \cite{zhu2021deep} where the context is related with the rewards using neural networks. The details of SAU-Sampling will be discussed in following sections.

\subsection{Multi-Agent Multi-Armed Bandits (MA-MABs)}
Multi-agent Multi-Armed Bandits is the multi-agent variant of MABs in which $N$ agents pull their $j^{th}$ arm and each $m^{th}$ agent will receive a reward drawn from their distribution $D_{m,j}$ with an unknown mean $\mu_{m,j}^*$ \cite{NEURIPS2021_c96ebeee}. MA-MABs can be modeled as centralized or distributed. In centralized settings the agents' actions are taken by a centralized controller and in distributed settings each agent will independently choose their own actions. Distributed decision-making settings scale more effectively \cite{Landgren2019} and naturally deals easily with large $K$ set of arms when compared with centralized settings that suffers of $K$ arms' cardinality explosion. Finally, the total regret can be defined as:
\begin{myequation}
    R_T = \sum_{t=1}^{T}\sum_{m=1}^{N} \mathbb{E}[(\text{max}_j\mu_{m,j}^* - \mu_{m,j}^*)] 
\end{myequation}
In this work, we consider two main approaches: distributed non-cooperative and cooperative MA-MABs with adversarial rewards.

\subsection{Deep Transfer Reinforcement Learning}
Transfer learning or knowledge transfer techniques improve learning time efficiency by utilizing prior knowledge. Typically, this is done by extracting the knowledge from one or diverse source tasks and then applying such knowledge in a target task \cite{Pan10}. If the tasks are related in nature and the target task benefits positively with the acquired knowledge from the source, then it is called inductive transfer learning \cite{Scott2018}. This type of learning is not uncommon and it is used by the human brain on a daily basis.  However, a phenomena called negative transfer can occur, if after knowledge transfer, the target task performance is negatively affected \cite{Zhuang2021}. 

In the realm of transfer learning we can find Deep Transfer Learning (DTL). DTL is a subset of transfer learning that studies how to utilize knowledge in deep neural networks. In the context of classification/prediction tasks, large amount of data is required to properly train the model of interest \cite{Vu2020}. In many practical applications where training time is essential to respond to new domains \cite{Elsayed2021}, retraining using large amount of data is not always feasible and possibly catastrophic in terms of performance. ``What to transfer" corresponds to one of the main research topics in transfer learning. Specifically, in the case of deep transfer learning four categories have been broadly identified: instances-based transfer, where data instances from a source task are utilized; mapping-based transfer, where a mapping of two tasks is used on a new target task; network-based transfer, where the network pre-trained model is transferred to the target task; and adversarial-based transfer, where an adversarial model is employed to find which features from diverse source tasks can be transferred to the target task\cite{Tan2018}. 

\begin{figure}[h]
\center
  \includegraphics[scale=0.32]{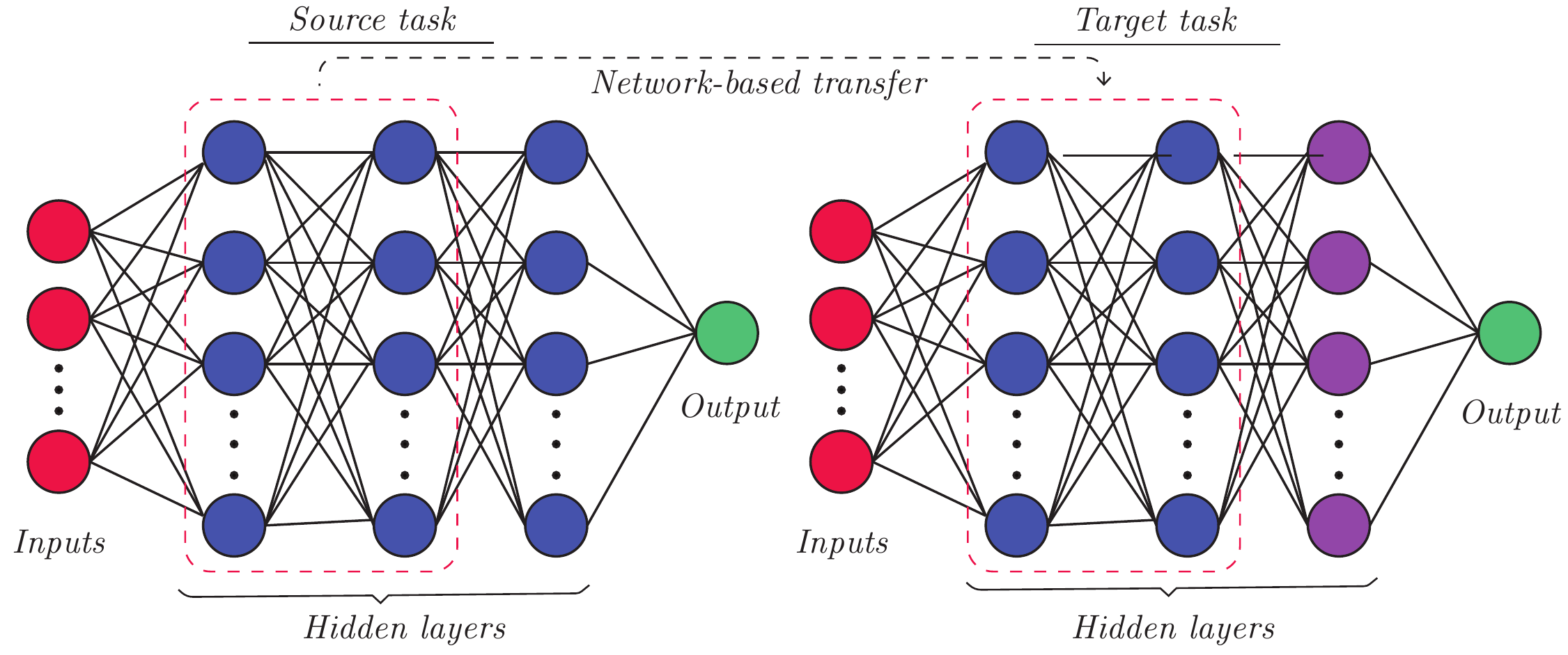}
  \setlength{\belowcaptionskip}{-5pt}
  \caption{Network-based transfer learning: the neural network source task's hidden layers are reutilized in the target network}
  \label{transfer_network}
\end{figure}

In this work, we utilize the DTL form called network-based transfer learning to adapt efficiently TP and CCA parameters in dynamic scenarios. An example of network-based transfer learning technique is presented in Fig. \ref{transfer_network}. Such technique is utilized in deep transfer reinforcement learning as part of a transfer learning type called policy transfer \cite{zhu2020transfer}. In particular, policy transfer takes a set of source policies $\pi_{S_1}, ..., \pi_{S_K}$ that are trained on a set of source tasks and uses them in a target policy $\pi_{T}$ in a way that is able to leverage the former knowledge from the source policies to learn its own. More specifically, the weights and biases that comprise each of the hidden layers of the source policies are the elements transferred to the target polices. Note that in practice policies are modeled as neural networks. 

In this paper, we take advantage of the design of a contextual multi-armed bandit presented in \cite{zhu2021deep} and apply policy transfer to improve the agent's SR adaptability in dynamic environments. The results and observations of applying DTRL are discussed in section \ref{adaptiveSR}. In the next section, we will discuss the details of the system model and present an analysis on reducing the cardinality of the action space in the proposed SR problem formulation. 

\section{System model and Problem Formulation} \label{Section4}

\begin{table}
\caption{Notations}
\centering
\label{param-def}
\begin{tabular}{p{1.74cm}|p{6.24cm}}

\textbf{Notation}       & \textbf{Definition}\\ \hline
$s$ and $\mathcal{S}$   & Index and set of stations, \\
$m$ and $\mathcal{M}$   & Index and set and the number of \noteblue{APs} \sout{RUs}, \\
 $x^{|\mathcal{S}|}$ and $c^{|\mathcal{M}|}$      & Stations' positions and AP's positions  \\
 \hline
$P_{cs}^{m}$                & CCA threshold of $m^{th}$ AP, \\
$P_{tx}^{m}$                & Transmission Power of $m^{th}$ AP, \\
$R_{s}^{m}$                 & Throughput of $s^{th}$ STA of $m^{th}$ AP, \\
$R_{s,A}^{m}$                & Achievable throughput of $s^{th}$ STA of $m^{th}$ AP,  \\
$D_{s}^{m}$                & Adaptive data link rate of $s^{th}$ STA of $m^{th}$ AP  \\
\hline
$P_{IDLE}^{m}$  & Probability of a STA is idle in a BSS, \\
$P_{SUCC,s}^{m}$               & Probability of succesful transmission by station $s^{th}$ STA to the $m^{th}$ AP,\\
$\phi_s^m $                  & Probability of $s^{th}$ STA be transmitting to the $m^{th}$ AP,  \\
$\xi_{CCA}$              & Binary function, $\xi_{CCA} = 1$ if signal is bellow the CCA threshold $P_{cs}$,  \\
$\xi_{ED}$              & Binary function, $\xi_{ED} = 1$ if signal is bellow the Energy Detection (ED) threshold $P_{ed}$,  \\
$\xi_{STA}$              & Binary function, $\xi_{STA} = 1$ if throug is bellow the Energy Detection (ED) threshold $P_{ed}$,  \\
 $E(T_{g,s}^m)$ and  $E(I_{g,s}^m)$ & Expected length of general time-slot and expected information transmitted by the $s^{th}$ STA of $m^{th}$ AP,  \\
 $T_{TXOP}$ and $T_{EDCA}$  & Packet transmission duration  and time required for a successful Enhanced Distributed Channel Access (EDCA) transmission,  \\
 $\Bar{P}^{fair}$ and $\Bar{U}$ & Average linear product-based network fairness and average station starvation,  \\
  $\omega$, $g_s^m$  and $\sigma^2$  & Fraction of  $R_{s,A}^{m}$ in which STAs are consider in starvation, the channel power gain and the power noise. \\
 
 \hline
$P_{tx}^m$ and $P_{tx}^r$           & The transmission power at the $m^{th}$ transmitter (AP) and the received signal strength at the $r^{th}$ receiver,  \\
$d_{m,r}$ and $\theta$           & Distance between the $m^{th}$ transmitter and $r^{th}$ receiver and path loss exponent,   \\
$\mathcal{F}_m^{+}$ and $\mathcal{F}_m^{-}$         & Subset of interferers and non-AP interferers,   \\
$\gamma_{m,r}$, $C_{m,r}$  and $C_T$      & Worst-case SINR and Shannon's maximum capacity of $m^{th}$ transmitter and $r^{th}$ receiver and cumulative maximum network capacity. \\

\hline
\end{tabular}
\label{notations}
\end{table}

In this work, we consider an infrastructure mode Wi-Fi 802.11ax network $\mathcal{N}$ with $N = |\mathcal{S}| + |\mathcal{M}|$ nodes where $\mathcal{S}$ is the set of stations with $\{\bm{x}^1, \bm{x}^2,..., \bm{x}^{|\mathcal{S}|}\} \in \mathbb{R}^2$ positions and $\mathcal{M}$ is the set of APs with $\{\bm{c}^1, \bm{c}^2,..., \bm{c}^{|\mathcal{M}|}\} \in \mathbb{R}^2$ positions. We can assume that $|\mathcal{M}|$ APs positions correspond to cluster centers and the stations will attach to their closest AP. In addition, the list of notations utilized in this work can be found in Table \ref{notations}.

In this paper, we improve SR via maximization of the linear product-based fairness and minimization of the number of stations under starvation by configuring TP and CCA parameters.
\begin{subequations}\label{opt-Verbal-CCmanagement}
	\begin{align}
	\label{opt-Verbal}
	& \textbf{Max}
	&& \begin{pmatrix} 
	    \text{fairness} \\
	    \text{avg. station starvation complement}
	    	    \end{pmatrix}
	\\
	\label{opt-Verbal2}
	& \textbf{s.t.}
	&& \text{Throughput}
	\\
	&\textbf{var.}
	&& \text{Transmission power and CCA threshold selection}
	\end{align}
\end{subequations}

Let's define the probability of an STA being idle in a BSS as:
\begin{align}
    P_{IDLE}^{m} = \prod_{s \in \mathcal{S}} \phi_s^{m'} &&\forall m\in \mathcal{M}.
\end{align}
where $\phi_s^m \in [0,1]$ is the probability of an STA transmitting to the $m^{th}$ AP. 
In addition, we proceed to define the probability in which an STA will successfully transmit a packet as:
\begin{align}
    P_{SUCC,s}^{m} = \phi_s^m\xi_{CCA}^{m}(\cdot)\xi_{ED}^{m}(\cdot)\prod_{s'\in\mathcal{S}, s'\neq s}^{\mathcal{S}}\phi_s^{m'} &&\forall m\in \mathcal{M}.
\end{align}
where $\xi_{CCA}(\cdot) = 1$ if the sensed signal of a packet sent by the $s^{th}$ STA is below the CCA threshold ($P_{cs}$), otherwise becomes zero. Here, $\xi_{ED}(\cdot) = 1$ if the sensed signal of packet sent by the $s^{th}$ STA is below the Energy Detection (ED) threshold ($P_{ed}$), otherwise becomes zero. Additionally, we consider $P_{cs} = P_{ed}$ to simplify our analysis.
As indicated by \cite{Derakhshani2018} the expected length of the general time-slot $\mathbb{E}(T_g)$ and the expected information transmitted by the $s^{th}$ STA to $m^{th}$ AP  $\mathbb{E}(I_g)$ can be expressed as:
\begin{align}
    E(T_{g,s}^m) = \delta P_{IDLE}^{m} + P_{IDLE}^{m'}T &&\forall m\in \mathcal{M}.
\end{align}
\begin{align}
    E(I_{g,s}^m) = P_{SUCC,s}^{m}D_s^m T_{TXOP} &&\forall m\in \mathcal{M}, s\in \mathcal{S}.
\end{align}
where $D_s^m$ corresponds to the link data rate, $T_{EDCA}$ corresponds to the time required for a successful Enhanced Distributed Channel Access (EDCA) transmission, $T_{TXOP}$ is the transmission duration  and $\delta$ the duration of an idle time slot. The link data rate will adaptively depend on SNR \cite{Holland2001} and mapped based on SNR/BER curves\cite{Riley2010}. The received SNR can be defined as $P_{tx}^m g_s^m\//\sigma^2$ where $P_{tx}$ is the transmission power, $g_s^m$ the channel power gain and $\sigma^2$ the power noise. 

Finally, the throughput of the $s^{th}$ station attached to the $m^{th}$ AP can be defined as: 
\begin{align}
    \label{thr_eq}
    R_s^m = \frac{E(I_{g,s}^m)}{E(T_{g,s}^m)} = \frac{P_{SUCC,s}^{m} D_s^m T_{TXOP}}{P_{IDLE}^{m}\delta  + P_{IDLE}^{m'} T_{EDCA}  },
\end{align}
Additionally, let's define the average linear product-based network fairness and average station starvation in a distributed setting:  
\begin{align}
\Bar{P}^{fair}(t) = \frac{1}{|\mathcal{M}|}\sum_{m \in \mathcal{M}}\prod_{s \in \mathcal{S}} \frac{R_s^m}{R_{m,A}^s},
\end{align}
\begin{align}
\Bar{U}(t) = \frac{1}{|\mathcal{M}|}\sum_{m \in \mathcal{M}}\frac{1}{|\mathcal{S}|}\sum_{s \in \mathcal{S}} \xi_{STA}(R_s^m > \omega R_{m,A}^s),
\end{align}
where $R_{m,A}^s$ is the achievable throughput of the $s^{th}$ station attached to the $m^{th}$ AP. Additionally, $\xi_{STA} = 1$ if $s^{th}$ station's throughput is greater than a fraction $\omega \in (0,1]$ of the achievable throughput, otherwise becomes zero in which case the station is considered in starvation.   
The considered problem is a multi-objective problem and can be addressed with the weighted sum approach. Thus, in each time step, the problem can be formulated as follows:  
\begin{problem}\label{problem_opt}
\begin{align}
 &\underset{\mathbf{P_{tx}},\mathbf{P_{cs}}}{\operatorname{max}}\;A_1 \Bar{P}^{fair}(t)+A_2 (1-\Bar{U}(t))\\
 &\text{s.t.}\nonumber\\
 &\text{\eqref{thr_eq}},\\
 &P_{tx}^{m}\in [P_{tx}^{min}, P_{tx}^{max}], P_{cs}^{m} \in [P_{cs}^{min},P_{cs}^{max}] &&\forall m \in \mathcal{M}
\end{align}
\end{problem}

Due the dynamic nature of the scenario, the transmission probabilities of the STAs $\phi_s^m$ are not directly controllable and require an additional step to map them to EDCA parameters \cite{Derakhshani2018}. Instead, we simplify our analysis by utilizing a network simulator to model such dynamics and propose to solve the previous linear programming (LP) problem using a MA-MAB solution as described in section \ref{Section5}.

\subsection{Optimal action set via worst-case interference} \label{worst-case}
Wi-Fi typical scenarios consist in APs and stations distributed non-uniformly. Contrary to the analysis presented in \cite{Kim2006} we aim obtaining an optimal subset of TP and CCA threshold values to further reduce the action space size in SR problems. In this analysis, we only consider the Carrier Sense (CS) threshold term as form of the CCA threshold. 

First, let's consider the worst-case interference scenario in a $N >2$ arrangement. For the sake of simplicity we use the path-loss radio propagation model: 
\begin{myequation}
    P_{rx}^{r} = \frac{P_{tx}^{m}}{{d_{m,r}}^{\theta}},
\end{myequation}
where $P_{tx}^{m}$ and $P_{rx}^{r}$ are the TP at the $m^{th}$ transmitter (AP) and the received signal strength at the $r^{th}$ receiver, respectively. In addition, $d_{m,r}$ is the distance between the transmitter and receiver. Finally, $\theta \in [2,4]$ corresponds to the path loss exponent. Thus, from the perspective of $m^{th}$ AP the worst-case interference $I_{m}$ is defined as:
\begin{myequation}
    I_m = \sum_{v \in \mathcal{F}_m^{+}} \frac{P_{tx}^{v}}{{X^{(m,v)}}^\theta} + P_{tx}^{sta}\sum_{w\in \mathcal{F}_m^{-}} \frac{1}{{X^{(m,w)}}^\theta}, 
\label{interference}
\end{myequation}

where $\mathcal{F}_m^{+}$ is the subset of interferers $|\mathcal{F}_m^{+}|=|\mathcal{M}|-1$,  corresponding to APs interfering with the $m^{th}$ AP and $\mathcal{F}_m^{-}$ the subset of non-AP interferers $|\mathcal{F}_m^{-}| = |\mathcal{S}|$, corresponding to the stations interfering with the $m^{th}$ AP. Furthermore, $P_{tx}^{v}$ is the TP of the  $v^{th}$ interferer and $P_{tx}^{sta}$ is a constant corresponding to the fixed power assigned to all the stations based on the fact that typically stations are not capable to modify their TP. Additionally, $X^{(m,v)}$ and $X^{(m,w)}$ corresponds to the distance from the  $m^{th}$ AP to the $v^{th}$ AP interferer and $m^{th}$ AP to the $v^{th}$ station interferer, respectively. $X^{(m,.)}$ is calculated as follows:
\begin{myequation}
  X^{(m,.)} = \sqrt{(D_m+x_{m,.})^2 + {d_{m,r}^2 - 2(D_m+x_{m,.})d_{m,r}\cos \varsigma_{r,.}}},
  \label{distance}
\end{myequation}
where $(.)$ refers either to the AP or non-AP interferer, $D_m$ is the CCA threshold range of the $m^{th}$ AP, $\varsigma_{r,.}$ is the distance between the receiver to the interferer $(.)$ and  $x_{m,.}$ corresponds to the distance between any $(.)$ interferer and $D_m$.

The corresponding worst-case SINR $\gamma_{m,r}$ at the receiver is defined as: 
\begin{equation}
    \gamma_{m,r} = \frac{P_{tx}^{m}}{{d_{m,r}}^{\theta} (I_m + N_{0})},
\end{equation}
Let's assume that $N_0 << I_m$, thus the equation is reduced to: 
\begin{equation}
    \gamma_{m,r} = \frac{P_{tx}^{m}}{{d_{m,r}}^{\theta}I_m },
    \label{sinr_formula}
\end{equation}
Substituting equations (\ref{interference}) and (\ref{distance}) in (\ref{sinr_formula}) we obtain equation (\ref{p_final}):

\begin{strip}
    \begin{align}
   \gamma_{m,r}= 
    \frac{\frac{P_{tx}^{m}}{{d_{m,r}}^{\theta}}}{\sum_{v\in \mathcal{F}_m^{+}} \frac{P_{tx}^{m}}{({\sqrt{(D_m+x_{m,v})^2 + {d_{m,r}}^2 - 2(D_m+x_{m,v})d_{m,r}\cos \varsigma_{r,.}}})^\theta} + P_{tx}^{sta}\sum_{w\in\mathcal{F}_m^{-}} \frac{1}{({\sqrt{(D_m+x_{m,w})^2 + d_{m,r}^2 - 2(D_m+x_{m,w})d_{m,r}\cos \varsigma_{r,w}}})^\theta}}
    \label{p_final}
\end{align}
\end{strip}

The aforementioned equation describes  $\gamma_{m,r}$ in function of $D_m$ and $d_{m,r}$. Additionally, we substitute $D_m =\left(P_{tx}^{m}/T_{cs}^{m}\right)^{1/\theta}$ in equation (\ref{p_final}), obtaining:

\begin{myequation}
 \gamma_{m,r}=\frac{\frac{P_{tx}^{m}}{{d_{m,r}}^{\theta}}}{\sum_{v\in\mathcal{F}_{m}^+}\frac{P_{tx}^{m}}{\Gamma^{m} + P_{tx}^{sta}\sum_{w=1}^{K_{m}^-} \iota^{(m,w)}} },
\end{myequation}

where, 
\[ \scalebox{.7}{$\Gamma^{m} = \left({\sqrt{\left[\left(\frac{P_{tx}^{m}}{T_{cs}^{m}}\right)^\frac{1}{\theta}+x_{m,v}\right]^2 + d_{m,r}^2 - 2\left[\left(\frac{P_{tx}^{m}}{T_{cs}^{m}}\right)^\frac{1}{\theta}+x_{m,v}\right]d_{m,r}\cos \varsigma_{r,v}}}\right)^\theta$},\]

$\iota^{(m,w)} = \frac{1}{{(\sqrt{(\Omega_{sta}+x_{m,w})^2 + d_{m,r}^2 - 2((\Omega_{sta}+x_{m,w})d_{m,r}\cos \varsigma_{r,w}}})^\theta}$ and
$\Omega_{sta} = \left(\frac{P_{tx}^{sta}}{T_{cs}^{sta}}\right)^\frac{1}{\theta}$. 

Now, we proceed to define the maximum channel capacity in terms of TP and Carrier Sense (CS) threshold ($T_{cs}$). Given a certain value of SINR, the Shannon maximum capacity is expressed as:
\begin{myequation}
    C_{m,r} = W\log_2(1 + \gamma_{m,r}),
\label{capacity}
\end{myequation}

where $W$ is the channel bandwidth in Hz. Then, the cumulative  maximum network capacity can be calculated as:

\begin{myequation}
    C_T = \sum_{m=1}^{|\mathcal{M}|-1}\sum_{r=1}^{N} C_{m,r},
\end{myequation}
%where $N$ is the total number of transmitters considering the worst-case scenario which in the case of a downlink scenario matches with the number of APs in the network.

\begin{figure}[h]
\center
  \includegraphics[scale=0.50]{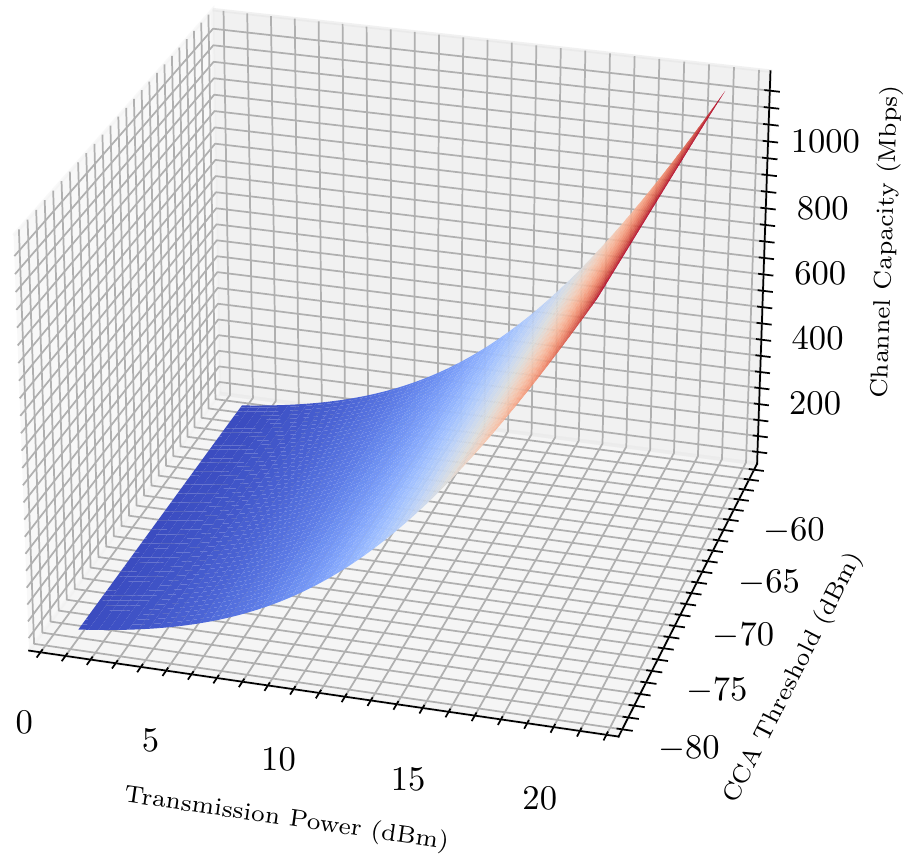}
  \setlength{\belowcaptionskip}{-5pt}
  \caption{Network capacity as a function of TP and CS threshold.}
  \label{network_capacity}
\end{figure}

In figure \ref{network_capacity}, it is shown a graph of the network maximum capacity as a function of TP and CS threshold. As observed, the network capacity achieves its higher values when a combination of high TP and low CS threshold is utilized. \iffalse That allows to select a small set from the action space in order to improve exploration time and consequently, convergence time.\fi Note that, prior knowledge of the locations are required.

\section{Proposed Multi-Agent Multi-armed bandit algorithms}\label{Section5}
In this section, we present the action space, context definition and reward function for the MA-MAB algorithms utilized in this work.

\subsection{Action space} \label{action_space}
The action space corresponds to the number of combinations of $P_{cs}$ and $P_{tx}$ which in the context of MABs translates to the number of arms for each MAB agent. The action space is defined as:
\begin{myequation}
    A_{cs} = \{P_{cs}^{min}, P_{cs}^{min} + \frac{P_{cs}^{max} - P_{cs}^{min}}{L_{cs}-1},..., P_{cs}^{max} \},
\end{myequation}

\begin{myequation}
    A_{tx} = \{P_{tx}^{min}, P_{tx}^{min} + \frac{P_{tx}^{max} - P_{tx}^{min}}{L_{tx}-1},..., P_{cs}^{max}\},
\end{myequation}
where $P_{cs}^{min}$, $P_{cs}^{max}$ and $P_{tx}^{min}$, $P_{tx}^{max}$ are the minimum and maximum values of CCA threshold and TP values, respectively. $L_{cs}$ and $L_{tx}$ corresponds to the number of levels to be discretized the CCA threshold and TP values, respectively. Finally, the number of arms corresponding to the action space for the $m^{th}$ agent $K_{m}^{AP}$ is $|A_{cs}^{m}| \cdot |A_{tx}^{m}|$. 

\subsection{Reward function in distributed non-cooperative settings}
The reward is defined following the optimization problem \ref{problem_opt}. The reward resembles the reward presented in \cite{Bardou2021} which includes a linear product-based fairness and station's starvation term  \cite{app112211074,Bardou2021} but defined in a distributed manner. A station is considered to be on starvation when its performance is bellow to a predefined percentage of its theoretical achievable throughput. The reward is defined as: 

\begin{equation}
    \resizebox{0.9\columnwidth}{!}{$r_{m}^{AP} = \frac{|\Psi_m^{AP}|\prod_{j\in \Psi_m^{AP}} \frac{R_m^s}{\omega R_{m,A}^s} + |N_m^{AP} \setminus \Psi_m^{AP}|(N_m^{AP} + \prod_{j\in N_m^{AP} \setminus \Psi_m^{AP}} \frac{R_m^s}{R_{m,A}^s})}{N_m^{AP}(N_m^{AP} + 1 )}$}, 
\end{equation}

where $\Psi_m^{AP}$ is the set of starving stations attached to the $m^{th}$ AP , $N_m^{AP}$ the set of stations attached to the $m^{th}$ AP. We can also observe, that $r_{m}^{AP} \propto C_{m,r}$ as defined in Eq. \ref{capacity}. 

In the next subsection, we present the definition of the context considered in our MA-CMAB solution.

\subsection{Distributed Sample Average Uncertainty-Sampling MA-CMAB }
In \cite{zhu2021deep}, the authors present an efficient contextual multi-arm bandit based on a ``frequentist approach'' to compute the uncertainty instead of using bayesian solutions as Thompson Sampling. The frequentist approach consist in measuring the uncertainty of the action-values based on the sample average rewards just computed instead of relaying on the posterior distribution given the past rewards. In this work, we present multi-agent cooperative and not cooperative  variants of the previously mentioned RL algorithm.   

In our problem, the context is comprised only by the APs’ local observations:
\begin{enumerate}
	\item Number of starving stations, $|\Psi_m^{AP}|$ where $m$ corresponds to the $m^{th}$ AP under $\omega$ fraction of their attainable throughput during the $t$ episode. 
	\item Average RSSI, $\overline{S}_m^{AP}$ where $m$ is the $m^{th}$ AP during the $t$ episode.
	\item Average Noise, $\overline{\Upsilon}_m^{AP}$ where $m$ denotes the $m^{th}$ AP during the $t$ episode.
\end{enumerate}

Additionally the context is normalized as follows: 
\begin{myequation}
    \psi_m^{AP} = |\Psi_m^{AP}|/N_m^{AP},
\end{myequation}

\begin{myequation}
    s_m^{AP} =\begin{cases} 
                    0,  & -50 \text{ dBm} \leq \overline{S}_m^{AP} \leq -60 \text{ dBm},   \\
                    0.25, &  -60 \text{ dBm} \leq \overline{S}_m^{AP} \leq -70 \text{ dBm},\\
                    0.5, &  -70 \text{ dBm} \leq \overline{S}_m^{AP} \leq -80 \text{ dBm}, \\
                    0.75, & -80 \text{ dBm} \leq \overline{S}_m^{AP} \leq -90 \text{ dBm}, \\
                    1, & -90 \text{ dBm} \geq \overline{S}_m^{AP} 
       \end{cases}\\
\end{myequation}

\begin{myequation}
    \hat{\Upsilon}_m^{AP} = \overline{\Upsilon}_m^{AP}/100,
\end{myequation}

The multi-agent SAU-Sampling algorithm in its non-cooperative version (SAU-NonCoop) is described in Algorithm \ref{sau-sampling}.The algorithm starts by initializing action-value functions $\mu(\bm{x}_m|\bm{\hat{\theta}}_{m})$ as a deep neural networks  and the exploration parameters $J_{m,a}^2$ and $n_{m,a}$ for each $m^{th}$ AP. $n_{m,a}$ correspond to the number of times action $a$ was selected in the $m^{th}$ AP and $J_{m,a}^2$ is defined as an exploration bonus. In each environment step (Algorithm \ref{sau-sampling}, \texttt{line 2}), each agent will observe their local context and compute the selected arm given the reward prediction. In (Algorithm \ref{sau-sampling}, \texttt{line 11}) each CMAB agent will update $\bm{\hat{\theta}}_{m,a}$ using stochastic gradient descent on the loss between the predicted reward and the real observed reward. Finally, the exploration parameters are accordingly updated given the the prediction error as depicted in (Algorithm \ref{sau-sampling}, \texttt{line 12}). 

\normalem %%%% disable auto underline
\setlength{\textfloatsep}{0pt}%
\begin{algorithm}
\algsetup{linenosize=\small}
 \scriptsize

\textbf{Initialize} network $\bm{\hat{\theta}}_{m,a}$, exploration parameters $J_{m,a}^2(t=0) = 1$ and $n_{m,a}(t=0) = 0$ for all actions $a \in K_m$. 

\For{environment step $t\gets1$ \textbf{to} $T$}{ 
    \For{agent $m$} {
        Observe context ${\bm{x}_m(t)} = [\psi_m^{AP}(t), s_m^{AP}(t), \hat{\Upsilon}_m^{AP}(t)]$  \\
        \For{$a = 1,...,K_m$} {
            Calculate reward prediction $\hat{\mu}_{i,t}(t) = \mu(\bm{x}_m|\bm{\hat{\theta}}_{m})$ and $\tau_{m,a}^2(t) = J_{m,a}^2/n_{m,a}$\\ 
            $\tilde{\mu}_{m,a} \sim \mathcal{N}(\hat{\mu}_{m,a},n_{m,a}^{1-}\tau_{m,a}^2)$
        }
        Compute $a_{m}(t) = \text{\texttt{argmax}}_a(\{\tilde{\mu}_{m,a}(t)\}_{a \in K_m}\})$ if $t > K_m$, otherwise $a_{m}(t) \sim \mathcal{U}(0,K)$;\\
        Select action $a_m(t)$, observe reward $r_m^{AP}$;\\
        Update $\bm{\hat{\theta}}_{m,a}$ using SGD with gradients $\partial l_m/\partial \theta$ where $l_m= 0.5(r_m^{AP} - \hat{\mu}_{m,a}(t)) $ ; \\
        Update $J_{m,a}^2 \leftarrow J_{m,a}^2 + e_m^2$ using prediction error $e_m = r_m^{AP}(t) - \hat{\mu}_{m,a}(t)$  and $n_{m,a} \leftarrow n_{m,a} + 1$;
    }
}

\caption{SAU-Sampling MA-CMAB}
 \medskip
 \label{sau-sampling}
\end{algorithm}

\subsection{Cooperative Sample Average Uncertainty-Sampling MA-CMAB}
In this section we present a cooperative version of SAU-Sampling named SAU-Coop. Different from the non-cooperative version, the total reward $r_{m}^{C}$ considers the network Jain's fairness index in addition to their local reward $r_m^{AP}$ as:

\begin{myequation}
    r_{m}^{C} = r_{m}^{AP} + r_{\mathcal{J}},
\end{myequation}

where $r_{\mathcal{J}}$  as the overall network Jain's fairness index is defined as:
\begin{myequation}
    r_{\mathcal{J}} = \mathcal{J}(R_1,...,R_{N}) = \frac{(\sum_{m=1}^{|\mathcal{M}|} R_m )^2}{|\mathcal{M}|\cdot\sum_{m=1}^{|\mathcal{M}|}R_m^2},
\end{myequation}
where $R_m =\sum_{s=1}^{|\mathcal{S}_m|}R_s^m$ is the total throughput of all the $S_m$ stations of the $m^{th}$ AP.
\subsection{Reward-cooperative $\epsilon$-greedy MA-MAB }
In addition to the previous cooperative algorithm, we propose a cooperative approach based on the classical $\epsilon$-greedy strategy \cite{sutton2018reinforcement} that takes into account in the action's reward update a percentage of the average reward of other agents. This algorithm is described in Algorithm \ref{egreedy-coop}.
\normalem %%%% disable auto underline
\setlength{\textfloatsep}{0pt}%
\begin{algorithm}
	\caption{Reward-cooperative $\epsilon$-greedy MA-MAB}
\algsetup{linenosize=\small}
 \scriptsize
\textbf{Initialize} $\epsilon_m(t=0) = \epsilon_0$,  $Q_{m,a}(t=0)\leftarrow 0$, $N_{m,a}(t=0)\leftarrow 0$ and $\beta$. 

\For{environment step $t\gets1$ \textbf{to} $T$}{ 
\For{agent $m$} {
    Execute action $a_{m}(t)$: 
    $a_{m}(t) =\begin{cases} 
                    \text{\texttt{argmax}}_{k=1,...,K}  r_{k,i}(t) & \text{with probability} 1 - \epsilon_{m}(t) \\
                    k \sim \mathcal{U}(0,K) &  \text{o.w}
                \end{cases}$\\
    Calculate reward $r_m^{AP}(t)$ based on feedback of the environment\\
    Update $Q_{m,a}(t+1) = Q_{m,a}(t) + \frac{1}{N_{m}(t)}[(r_m^{AP} + \beta\cdot\frac{1}{M-1}\sum_{m=1}^{M^{-i}} r_m^{AP}) - Q_{m,a}(t)] $ \\
    Update $N_{m} \leftarrow N_{m}(t) + 1$;\\
    Update $\epsilon_{m} \leftarrow \frac{\epsilon_{m}(t)}{\sqrt{t}}$
 }
}
 \medskip
 \label{egreedy-coop}
 
\end{algorithm}

Finally, in the  next subsection we present the details of the the DTRL scheme to improve SR adaptation in dynamic environments. 
\subsection{Sample Average Uncertainty-Sampling MA-CMAB based Deep Transfer Reinforcement Learning }

Typically, RL agents learn their best policy based on the feedback received from the environment in a $T$ horizon time. However, in real-world scenarios the environment conditions can change in $T+1$ and thus, adapting to the updated environment is necessary\cite{Padakandla2021}. In such cases, the ``outdated'' agent's policy might not be optimal to address the new conditions efficiently. For instance, a modification on the stations' distribution over the APs can cause that the SR-related parameters chosen by the ``outdated'' agents' policy affect the network performance. 

\normalem %%%% disable auto underline
\setlength{\textfloatsep}{0pt}%
\begin{algorithm}
\algsetup{linenosize=\small}
\scriptsize
\textbf{Function} \textsc{Detect\_Singularity}($\mathcal{K}$) \tcp*[l]{returns True if anomaly is detected in network KPIs data $\mathcal{K}$ at time $t$,  and False otherwise.}
\textbf{Let} $\mathcal{L} = \{l | l \in \mathbb{N}, l>0\} $ the set of layers of model $\bm{\hat{\theta}}_{m,a}^l $ and $\mathcal{M} \subset \mathcal{L}$ the subset of layers to be transferred.
\text{\normalfont \textbf{Run} algorithm \textsc{SAU-Sampling MA-CMAB} \texttt{(Algorithm \ref{sau-sampling})}}
\While{environment step $t < T$} {
    
    \eIf{$\neg$\textsc{Detect\_Singularity}}
    {
    continue; 
    }
    {
       \text{\normalfont\textbf{Reset} exploration parameters $S_{m,a}^2, n_{m,a}$\;}
       
       \text{\normalfont \textbf{Reinitialize} weights $w$ and biases $b$ of the $l^{th}$ layer of $\bm{\hat{\theta}}_{m,a}^{l \not\in \mathcal{M}}$ via:} \\
        $\nu_{l} = \left(\sqrt{|\bm{\hat{\theta}}_{m,a}^{l\not\in \mathcal{M}}|}\right)^{-1}$ \;
        
        $\bm{\hat{\theta}}_{m,a}^{l \not\in \mathcal{M}} (w,b) \rightarrow w_{l} \sim \mathcal{U}(-\nu_l, \nu_l) , b_{l} \sim \mathcal{U}(-\nu_l, \nu_l) $\; 
        \text{\normalfont \textbf{Transfer }weights and biases via: } \\
         $\bm{\hat{\theta}}_{m,a}^{l \in \mathcal{M}} (w,b) \rightarrow \bm{\hat{\theta}}_{m,a}^{l \in \mathcal{M}'} (w,b)$\;
        
    }}

\caption{SAU-Sampling MA-MAB Transfer Learning}

\label{transfer_algo}

\end{algorithm}

To address the previous situation we propose two main solutions: \textbf{1.} If the agent detects a change in the environment indicated by a singularity, it will decide to correct its configuration via forgetting the policy already learnt (\textbf{forget}) or \textbf{2.} adapting the agent's policy to the new conditions via a transfer learning technique. A singularity is defined as a anomalous behavior of the KPIs of interest after the policy of the MAB agent has converged. In this work, we don't delve into how to detect a singularity and moreover, we assume the existence of an anomaly detector in our system \cite{10.1145/3444690}. In Algorithm \ref{transfer_algo}, we present the transfer learning algorithm depicting the second proposed solution. At $t=0$ each SAU-Sampling agent will reset their weights and biases and start learning as part of Algorithm \ref{sau-sampling}. At $t=S1$, where $S1$ corresponds to the time when an anomaly is detected and the transfer procedure is activated (Algorithm \ref{transfer_algo}, \texttt{line 7}). In our setup we transfer $l=2$ and reset $l=1$ (Algorithm \ref{transfer_algo}, \texttt{line 11}) , where $l$ corresponds to the layer of the neural network utilized in the SAU-Sampling agent. However, as indicated (Algorithm \ref{transfer_algo}, \texttt{line 13}), the transfer is not constrained to one layer but more generally to a set of layers. The set of transferred layers is considered as an hyperparameter to be tuned. The partial transfer of a model avoids negative transfer by giving the agent room to adapt to the new context since it mitigates model overfitting. 

\section{Performance Evaluation} \label{Section6}
\subsection{Simulation Setting}\label{AA}

We consider two scenarios in our simulations. The first one considers stationary users, meanwhile the second scenario considers mobile users to model dynamic scenarios (see section \ref{adaptiveSR}). In addition, stations and APs are two-antenna devices supporting up to two spatial streams in transmission and reception. In this work, we assume a frequency of 5 GHz with a 80 MHz channel bandwidth in a Line of Sight (LOS) setting. The propagation loss model is the Log Distance propagation loss model with a constant speed propagation delay. In addition, an adaptive rate data mode is considered with a UDP downlink traffic. We implement our proposed solutions  using ns-3 and also we use OpenAI Gym to interface between ns-3 and the MA-MAB solution\cite{Gawowicz2019}. In Table \ref{q_settings} and Table \ref{net_settings}  we present the learning hyperparameters and network settings parameters, respectively.
\begin{table} [ht]
  \centering
 \caption{Learning hyperparameters}
   \begin{threeparttable}
 \resizebox{\columnwidth}{!}{

\begin{tabular}{c c} 
\hline
\textbf{Parameter}&\textbf{Value} \\
\hline

$\epsilon$-greedy MAB & { Annealing $\epsilon$: $\sqrt{T}$} \\

Thompson Sampling MAB & { Prior distribution: Beta}\\

Upper Confidence Bound MAB & {Level of exploration, $c = 1$} \\

SAU-Sampling & { Number of hidden layers, $N_h=2$ } \\
        & {Number of neurons per hidden layer, $n_h=100$}\\
        & {Number of inputs, $N_m=3$ and number of outputs, $N_o = K$}\\
        & {Batch size, $B_s = 64$} \\
        & Optimizer :  {RMSProp (8e-3)} \\
        & Weight decay :  {5e-4} \\
        & Activation function : {ReLU}\\

\hline

Gym environment step time & { 0.05 s } \\

  \bottomrule
  \end{tabular}
  }

   \end{threeparttable}
  \label{q_settings}

\end{table}

\begin{table}
\caption{Network settings}
\begin{center}
\resizebox{\columnwidth}{!}{%
\begin{tabular}{c c} 
\hline
\textbf{Parameter}&\textbf{Value} \\
\hline

Number of APs & { 6 } \\
Number of Stations & {15}\\
Number of antennas (AP)  & {2} \\
Max Supported Tx Spatial Streams & { 2 } \\ %, 
Max Supported Tx Spatial Streams & {2} \\
Channel Number \footnotemark & { 1 } \\
Propagation Loss Model & { Log Distance Propagation Loss Model } \\
Wi-Fi standard & { 802.11 ax } \\
Frequency & { 5 GHz } \\
Channel Bandwidth & {80 MHz} \\
Traffic Model - UDP application & { $[0.011, 0.056, 0.11 \text{\cite{WILHELMI201926}}, 0.16]$ Gbps } \\
Maximum $\&$ minimum Transmission Power & $P_{tx}^{max}=21.0$ dBm $\&$ $P_{tx}^{min}=1.0$ dBm \\
Maximum $\&$ minimum CCA threshold & $P_{cs}^{max}=-62.0$ dbm $\&$ $P_{cs}^{min}=-82.0$ dbm\\
& $K_{cs} = 1$ and $K_{tx}= 1$\\

\hline
\end{tabular}
}
\label{net_settings}

\end{center}

\end{table}

%\vspace{-1cm}
\subsection{Reduced set of actions vs. all actions}\label{resultsA}
In subsection \ref{worst-case} we presented a mathematical analysis to obtain a reduced set of optimal actions with the goal of decreasing exploration time and consequently improving convergence time. As concluded in figure \ref{network_capacity}, high TP and low CCA threshold values maximize the network capacity in the simulation scenario under study. Therefore, we selected a fixed value of CCA threshold ($P_{cs}=-82.0$ dBm) and a reduced set of TP $P_{tx} \in \{15,16,17,18,19,20,21\}$ dBm and observed the performance against the full set of possible actions described in \ref{action_space}. 
\begin{figure}[t!]
\center
  \includegraphics[scale=0.38]{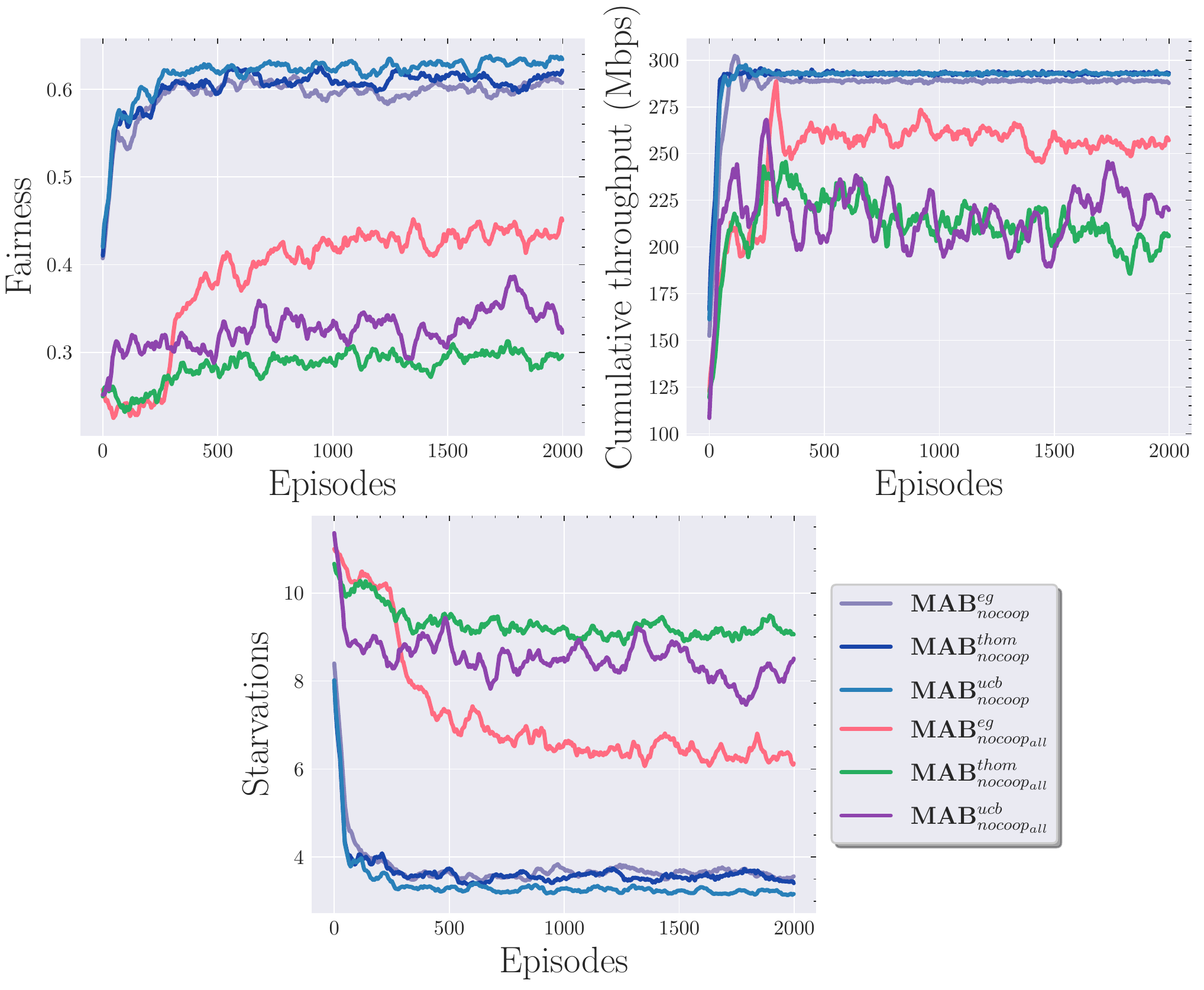}
  
  \caption{Convergence performance of $\epsilon$-greedy, UCB and Thompson Sampling MA-MABs under non-cooperative and distributed regimen. The subscript ``\textbf{all}" indicates the usage of the full set of actions.}
  \label{optimal_vs_all}
\end{figure}
In figure \ref{optimal_vs_all}, we present the convergence performance of three MA-MAB algorithms under UDP traffic of 0.056 Gbps in a non-cooperative and cooperative settings (indicated with subscripts ``\textbf{non-coop}" and ``\textbf{coop}", respectively ). The algorithms correspond to $\epsilon$-greedy ($MAB^{eg}$), UCB ($MAB^{ucb}$) and Thompson Sampling ($MAB^{thom}$) MA-MABs. For each algorithm, we plotted three convergence graphs in terms of fairness, cumulative throughput and station starvation representing the behavior when a reduced set of actions and the full action set (indicated with the subscript ``\textbf{all}") are used, respectively. For the case of the set of optimal actions, we can observe that the performance is similar with a slight improvement when utilizing MAB-Thompson Sampling. On the other hand, when utilizing the full action set the behavior shows a noticeable improvement with MAB $\epsilon$-greedy algorithm with respect the others. In \cite{NEURIPS2020_12d16adf}, the authors study the unreasonable behavior of greedy algorithms when $K$ is sufficiently large. They concluded that when $K$ increases above 27 arms, intelligent algorithms are affected greatly by the exploration stage. The former results validate ours based on the fact that $K   =|A_{cs}| \cdot |A_{tx}| = 21^2$. Finally, it can be noted that the impact of utilizing reduced optimal actions in terms of convergence time and KPI maximization. The set of optimal tasks allows to reduce the station starvation when compared with the best performer $MAB_{nocoop_{all}}^{eg}$ by an average of two starving users. However, in order to obtain such a set it is requires a prior knowledge of stations and APs geographical locations. In the following section we compare the results of  $\epsilon$-greedy MA-MAB and a default typical configuration without machine learning.  

\footnotetext{We assume all APs are configured to use 1 channel out of the available 11. This is a practical selection to create dense deployment scenarios.}

\begin{figure}[h]
\center
  \includegraphics[scale=0.37]{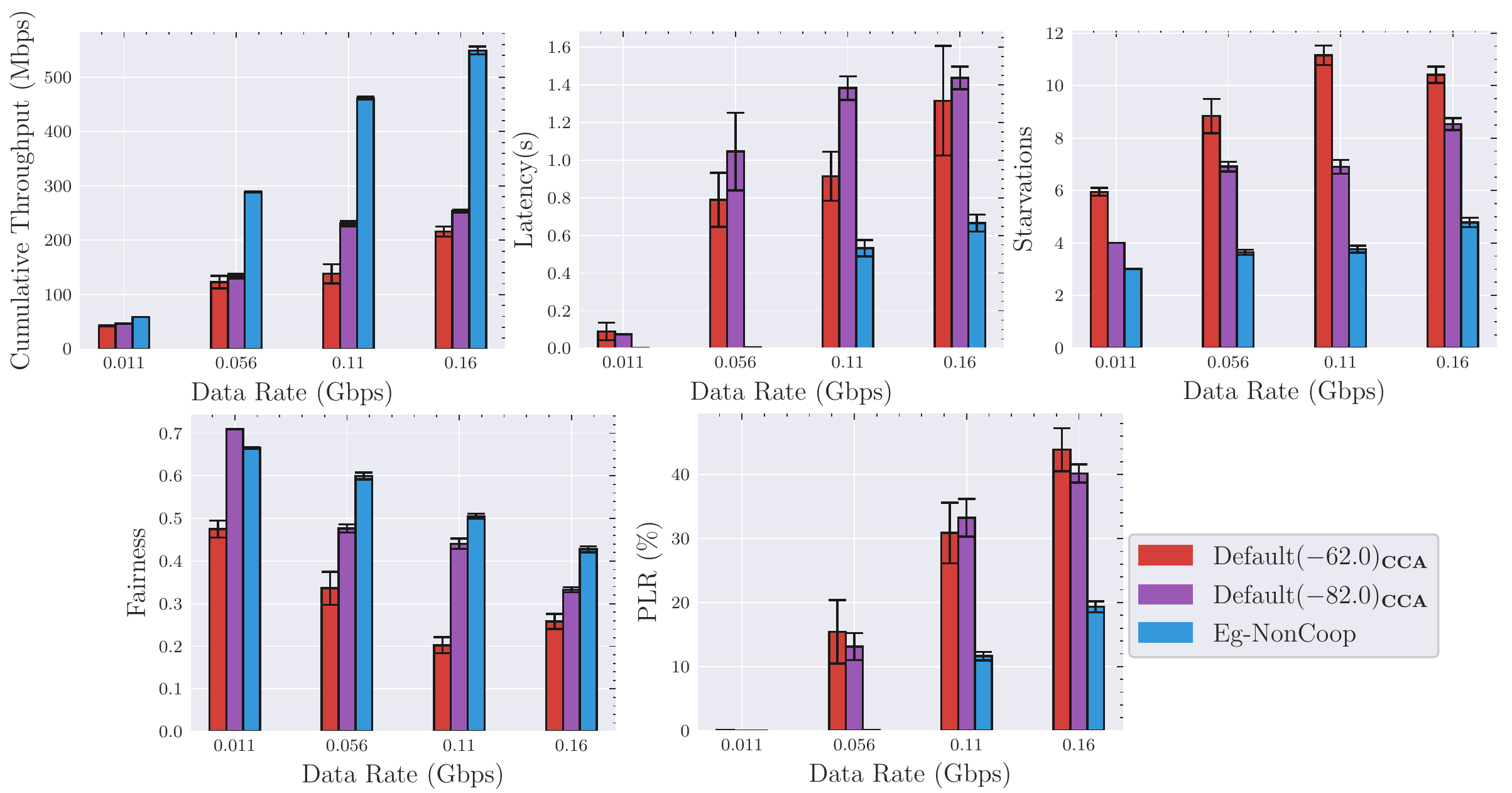}
  \caption{Performance results: $\epsilon$-greedy MAB w/ optimal set vs. default configuration with $P_{cs} \in \{-62.0, -82.0\} $ dBm. }
  \label{mabegreedy}
\end{figure}

\subsection{Distributed $\epsilon$-greedy MA-MAB vs. default configuration performance results}\label{dist_legacy}

In this subsection, we present the comparative results and advantages of utilizing a distributed intelligent solution such as MAB $\epsilon$-greedy over the default CCA threshold and TP configuration with no ML. In figure \ref{mabegreedy}, we show the performance under four different UDP data traffic regimes: $\{0.011, 0.056, 0.11, 0.16\}$ Gbps. We considered two typical configurations of CCA threshold: $-82.0$ dBm and $-62.0$ dBm. In both cases, the AP's TP is $16.0$ dBm. It can be observed that MAB $\epsilon$-greedy achieves a significant improvement over the default configuration ($P_{cs}=-82.0$ dBm) with an average gain over all the considered traffic of $44.4\%$ in terms of cumulative throughput, $70.9\%$ in terms of station starvation, $12.2\%$ in terms of fairness, $138.0\%$ in terms of latency and $94.5\%$ in terms of packet loss ratio (PLR), respectively. Additionally, a gain over the default configuration ($P_{cs}=-62.0$ dBm) with an average gain over all the considered traffics of $53.9\%$ in terms of cumulative throughput, $138.4\%$ in terms of station starvation, $43.0\%$ in terms of fairness, $84.0\%$ in terms of latency and $105.4\%$ in terms of packet loss ratio (PLR) is shown, respectively.

\begin{figure}[h]
\center
  \includegraphics[scale=0.38]{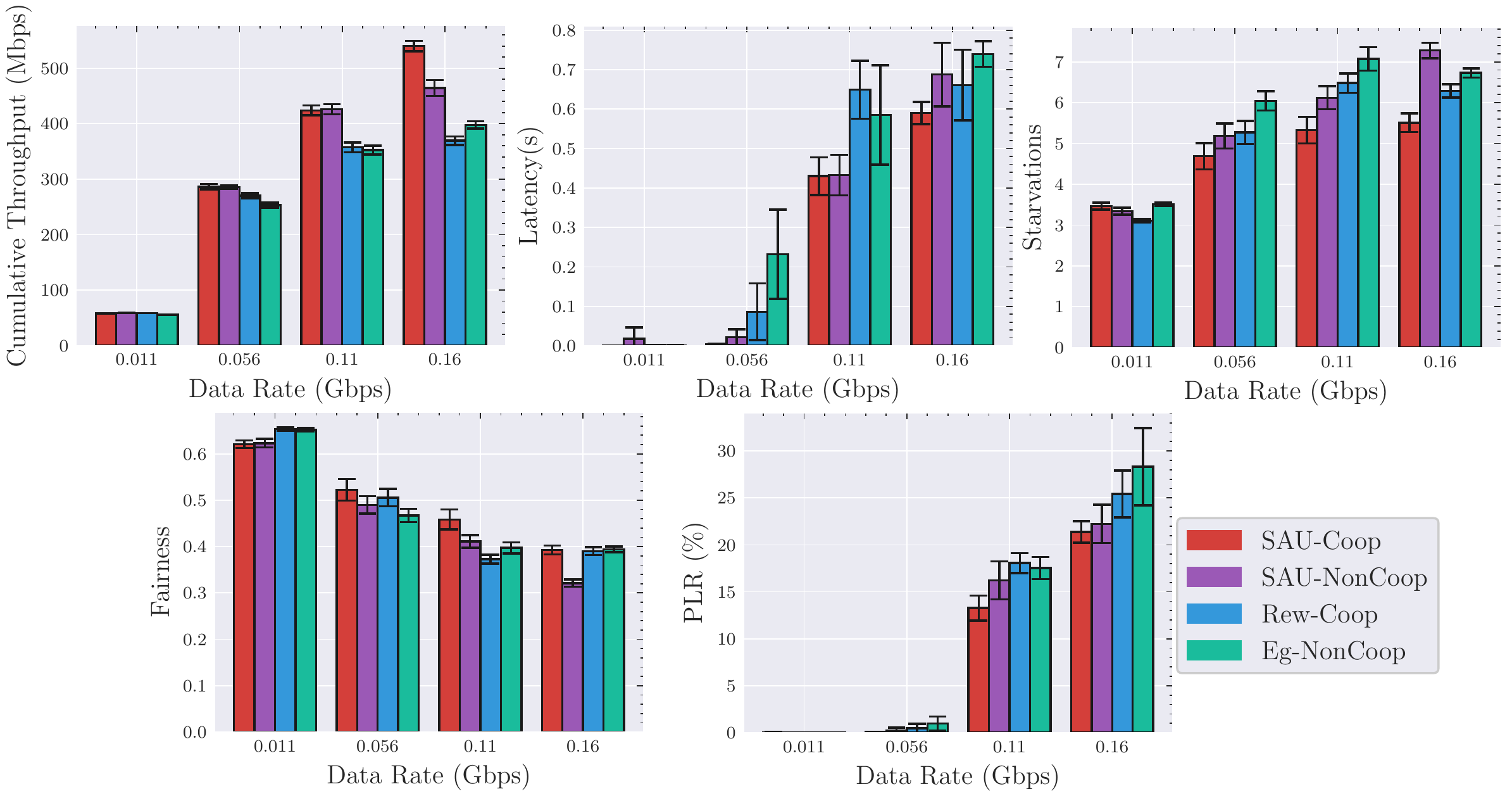}
  \caption{Performance results of cooperative algorithms: $\epsilon$-greedy MA-MAB (Rew-Coop), SAU-Sampling MA-CMAB (SAU-Coop) and non-cooperative versions of the previous algorithms SAU-NonCoop and Eg-NonCoop under full-set of actions.}
  \label{sau_results}
\end{figure}
\vspace{0em}

\subsection{Cooperation vs. non-cooperation performance results}\label{AA}

In the two past subsections we have shown the results considering the set of optimal actions. In this subsection we assume the non-existence of stations and APs location information and thus, we must rely on the full set of actions. In consequence, we investigate if cooperation can improve the KPIs of interest by utilizing the cooperative proposal of the MAB $\epsilon$-greedy algorithm (Rew-Coop) and the contextual SAU-Sampling algorithm (SAU-Coop). Additionally, we present two non-cooperative algorithms: SAU-NonCoop which corresponds to the non-cooperative version of the SAU-Sampling and Eg-NonCoop that refers to the MAB $\epsilon$-greedy algorithm utilized in the previous section. As observed in figure \ref{sau_results}, simulations show that SAU-Coop improves Eg-NonCoop over all the data traffic with an average of $14.7\%$ in terms of cumulative throughput, $21.3\%$ in terms of station starvation, $4.64\%$ in terms of network fairness, $36.7\%$ in terms of latency and  $32.5\%$ in terms of PLR. Similarly, the distributed version of SAU-Sampling presents a better performance over Eg-NonCoop, indicating that context is beneficial to solve the current optimization problem. Additionally, SAU-Coop presents a better performance over its non-cooperative version, specially when the data rate increases up to 0.16 Gbps where it is observed a gain of $14.1\%$ in terms of cumulative throughput, $32.1\%$ in terms of station starvation, $18.2\%$ in terms of network fairness, $16.5\%$ in terms of latency and  $4\%$ in terms of PLR. To sum up, cooperative approaches contribute positively to the improvement of SR in WiFi over non-cooperative approaches. In addition, in cases where cooperation is not possible it is advisable to utilize contextual multi-armed bandits over stateless multi-armed bandits.

\subsection{Deep Transfer Learning in Adaptive SR in Dynamic scenarios results}\label{adaptiveSR}

In order to model a dynamic scenario, we design a simulation where the users move across the  simulation area and attach to the AP that offers the best signal quality. Consequently, the user load in each AP will change and thus, the dynamics of the environment. We model this scenario with 3 APs and 15 users where the load will change twice throughout the simulation. As depicted in table \ref{dynamic_table} the user load of the $m^{th}$ AP denoted as $C_m$ will change in two instances in time: 3 and 6 minutes, respectively. %$U_{act}$ corresponds to the active users in the network.

\begin{table}[ht]
\caption{Dynamic scenario load distribution} % title of Table
\begin{center}
\centering % used for centering table
\begin{tabular}{c|cccccc|}

\cline{2-7} & \multicolumn{2}{c|}{\bfseries $\bm{t=0}$ min}  & \multicolumn{2}{c|}{\bfseries $\bm{t=3}$ min}          & \multicolumn{2}{c|}{\bfseries $\bm{t=6}$ min}  \\ \hline
\multicolumn{1}{|c|}{$C_1$}            & \multicolumn{2}{c|}{8}  & \multicolumn{2}{c|}{5}   & \multicolumn{2}{c|}{2}     \\ \hline
\multicolumn{1}{|c|}{$C_2$}            & \multicolumn{2}{c|}{5}  & \multicolumn{2}{c|}{5}   & \multicolumn{2}{c|}{2}      \\ \hline
\multicolumn{1}{|c|}{$C_3$}            & \multicolumn{2}{c|}{2}  & \multicolumn{2}{c|}{5}   & \multicolumn{2}{c|}{11}      \\ \hline
%\multicolumn{1}{|c|}{$U_{act}$}     & \multicolumn{2}{c|}{15} & \multicolumn{2}{c|}{15}  & \multicolumn{2}{c|}{15}    \\ \hline
\end{tabular}

\end{center}
\label{dynamic_table}
\end{table}
\vspace{0.0em}

\vspace{0em}

\begin{figure}[h]
\center
  \includegraphics[scale=0.4]{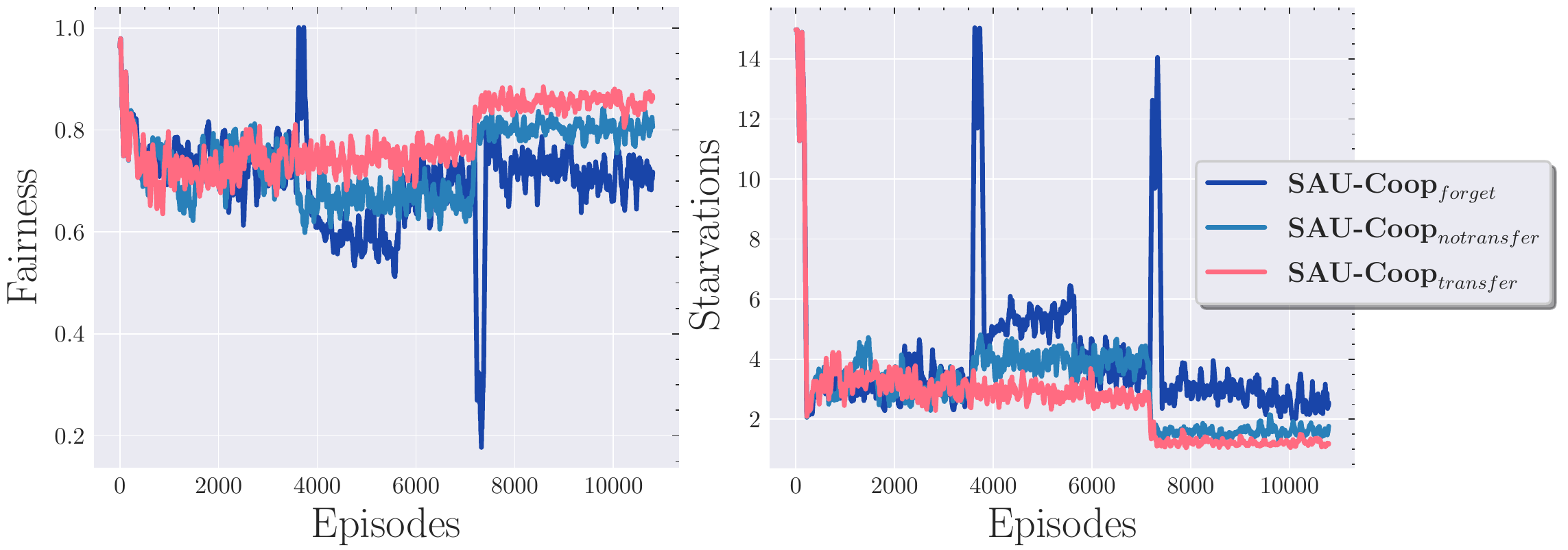}
  \setlength{\belowcaptionskip}{-5pt}
  \caption{Network response in terms of fairness and station starvation when utilizing the forget, full transfer and transfer strategies. }
  \label{transfer_forget_results}
\end{figure}

In figure \ref{transfer_forget_results} we present the network behavior in terms of fairness and station starvation under the scenario depicted by Table \ref{dynamic_table}. In addition to the two methods previously mentioned: \textbf{forget} and \textbf{transfer}, we present the performance of a third approach called \textbf{full transfer} where the full transfer of the model is considered. During the first interval ($0-3$min) the performance is similar in the three methods as expected. However, after the two changes on the network load, two singularities in each graph are visible in the fairness and starvation graphs. More specifically, the \textbf{forget} method experiences the worst behavior, with a $54.3\%$ and $11.7\%$ decrease when compared with the transfer method in terms of station starvation and fairness, respectively. The \textbf{forget} method shows some peaks at the moment of the singularities representing  $60\%$ of total of the users with a service drop; this behavior is inherently related to the agents' process of start learning again and cannot be avoided. From the quality of service perspective, a disturbance such as the one observed is highly non-preferable. Meanwhile, the \textbf{full transfer} method underperforms the \textbf{transfer} method with $18.7\%$ and $6\%$ decrease in the previously mentioned KPIs.  
Interestingly, it can be observed in the second interval under study ($3-6$min) the \textbf{forget} method is able to overperform at the end of the period the \textbf{full transfer} method. This is due to a negative transfer as a result of transferring the whole model. As observed, not only the partial transfer learning reduces considerably the peaks in performance of the \textbf{forget} method but also it is able to achieve better adaptation over the \textbf{full transfer} method. In all methods, the cumulative throughput is similar, however as observed in figure \ref{transfer_forget_results} station starvation and consequently, fairness are affected. 

\section{Conclusion} \label{Section8}

In this paper, we propose Machine Learning (ML)-based solutions to the Spatial Reuse (SR) problem in distributed Wi-Fi 802.11ax scenarios. We presented a solution to reduce the huge action space given the possible values of Transmission Power (TP) and Clear-Channel-Assessment (CCA) threshold values per Access Point (AP) and analysed its impact on diverse well-known distributed Multi-Agent Multi-Armed Bandit (MA-MAB) implementations. In distributed scenarios, we showed that $\epsilon$-greedy MA-MAB significantly improves the performance over typical configurations when the optimal actions are known. Moreover, the Contextual Multi-Agent Multi-Armed (MA-CMAB) named SAU-Sampling in the cooperative setting contributes positively to an increase in throughput and fairness and reduction of PLR when compared with no cooperation approaches. Under a dynamic scenarios, transfer learning benefits the SAU-Sampling algorithm to overcome the service drops for at least $60\%$ of the total of users when utilizing the forget method. Additionally, we obtained that partial transfer learning offers better results than the full transfer method. To conclude, the utilization of the cooperative version of the MA-CMAB to improve SR in WiFi scenarios is preferable since it outperforms the presented ML-based solutions and prevents service drops in dynamic environments via transfer learning.

\section{Acknowledgment }\label{Section9}
This research is supported by Mitacs Accelerate Program and NetExperience Inc.
%----------------------------------------------------------------------------------------
%	BIBLIOGRAPHY
%----------------------------------------------------------------------------------------

\bibliography{biblio.bib}{}
\bibliographystyle{IEEEtran}
% \printbibliography[title={Bibliography}] % Print the

\end{document}